# New neutron transmission supermirror remanent polarizer

## V.G. Syromyatnikov[1,2]

1 - *Petersburg Nuclear Physics Institute, National Research Center "Kurchatov Institute", Gatchina, Russia*
2 - *Department of Physics, Saint Petersburg State University, St. Petersburg, Russia*

## Abstract.

A new neutron transmission supermirror remanent polarizer is proposed. The polarizer is compact. It consists of two solid state polarizing parts. Parts of the polarizer have antiparallel magnetization of magnetic layers of supermirror structures. Both parts of the polarizer have magnetic remanence. This allows it to be used as a polarizer and analyzer in small magnetic fields. The polarizer can be used when working with beams of large cross-section and with a wide angular distribution. Assembly of the device is greatly simplified, because it is not required to bend a large stack of short plates along the radius, but only to compress of its. The width of the angular profile of the beam passed through the polarizer does not change! The polarizer has a high transmittance for the (-) spin component of the beam. The beam passed through the polarizer is highly polarized.

Properties of neutron polarizing *CoFe/TiZr* and *Fe/Si* remanent supermirrors (*m* = 2.0 and 2.5) are discussed.

A variant of effective use of the proposed polarizer with a spin-flipper for operation in saturating magnetic fields is proposed.

## 1. Introduction.

Recently, transmission neutron supermirror polarizers have become widespread in the neutron experiment using full neutron polarization analysis. One of the main advantages of such polarizers: the beam coming out of the polarizer has the same direction as the beam at the entrance. This allows you to quickly and easily rebuild the facility from measurements with polarized neutrons to measurements with non-polarized.

The neutron transmission supermirror polarizers are known: V-cavity [1-3] and solid state devices: transmission bender with collimator [4], S-shaped bender [5] and kink polarizer [6, 7].

The main parameters of these polarizers are performed in Supplement I.

The material of this paper was presented as oral contribution to European Conference on Neutron Scattering 2019 (ECNS 2019) on July 2$^{nd}$, 2019. A new neutron transmission solid state supermirror remanent polarizer is described in this paper, both for the polarization of large area beams with a wide angular distribution, as well as for the analysis of the polarization of such beams scattered on the sample.

Some examples of remanent supermirrors are presented in Supplement II.

# 2. Remanent polarizing neutron supermirrors.

As is known, the technique of magnetron sputtering is widely used to create polarizing neutron periodic and aperiodic (supermirror) multilayer structures. In the process of such sputtering, magnetic anisotropy with easy and hard magnetization axes occurs in the magnetic layers of the supermirror coating. The magnetization curve of such a multilayer structure (supermirror) along the easy axis is characterized by a high remanence, i.e. in the region of small fields $H \sim (10 - 20)$ Oe on both branches of the hysteresis curve, the magnetization of magnetic layers is high and close to its maximum value in absolute value. The shape of the hysteresis loop is close to a rectangle, as shown in Fig. 1. In an electromagnet, the transition from one magnetization to the opposite occurs quickly: by means of a current pulse.

If we are at point *A* of the hysteresis curve, the magnetization of the magnetic layers of the multilayer structure is oriented parallel to the guiding magnetic field and the reflected beam will be polarized along the field (see Fig. 1).

If we are at point B of the magnetization hysteresis curve of the magnetic layers, the supermirror is oriented antiparallel to the guiding field and the reflected beam will be polarized antiparallel to the field (see Fig. 1).

Using the remanence of polarizing supermirrors allows to work in the polarization analysis scheme without a spin-flipper in front of the sample.

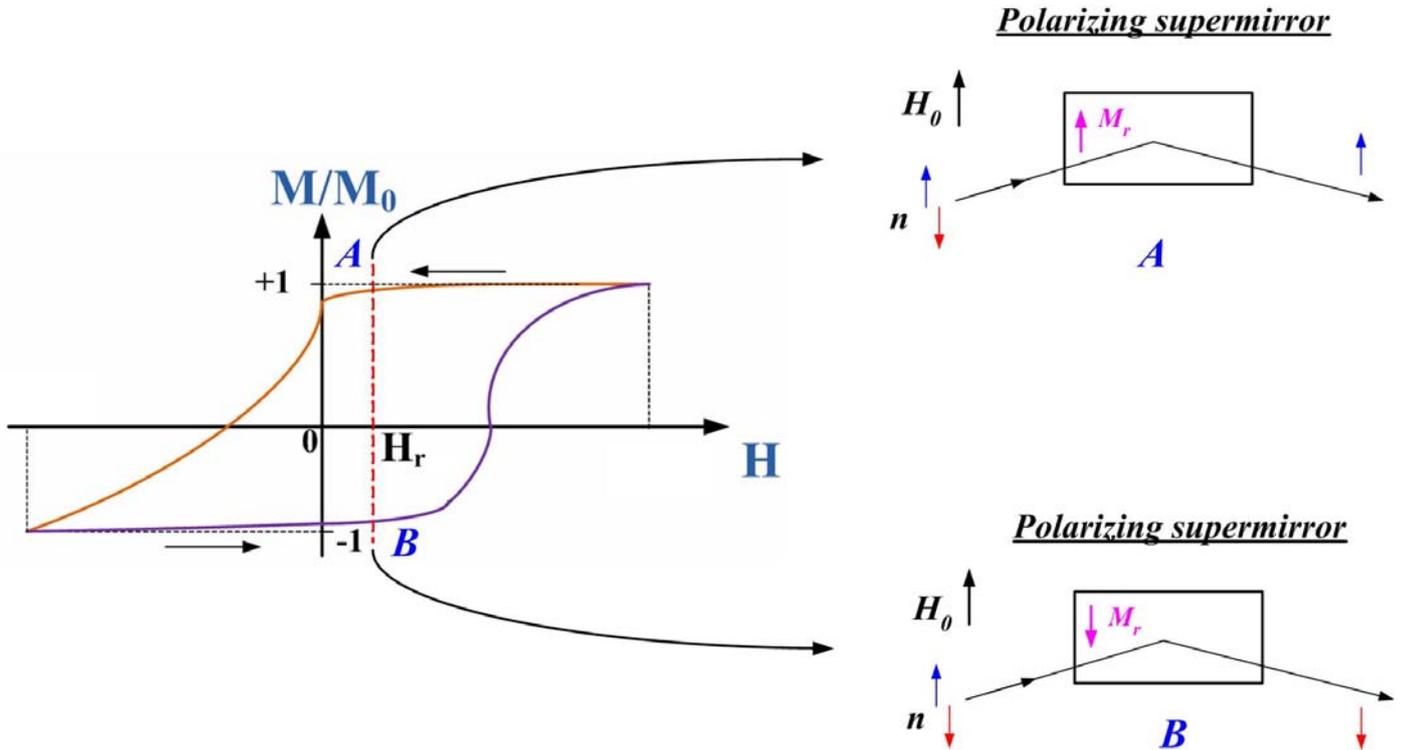

**Fig. 1.** Hysteresis loop for neutron remanent polarizing supermirrors.

The neutron remanent multilayer structures have been investigated in papers [8-11]. The main parameters of some remanent *CoFe/TiZr* and *Fe/Si* structures are performed in Supplement II.

# 3. Concept of new supermirror neutron polarizer

Consider the concept of a new neutron polarizer [12]. This is a supermirror solid state multichannel transmission polarizer. It uses the property of remanent of polarizing supermirrors. The polarizer consists of two compact multichannel solid state parts. In both parts, each channel is a plate of neutron-transparent material. The first part is a spin splitter, the second part is straight polarizing NG or collimator-polarizer (see Fig. 2). In part 1, the supermirror coating is sputtered directly on both sides of the plate. In the 2nd part, supermirror coatings are also sputtered on both sides of the plates. Antireflective absorbing sublayer is sputtered on top of them. The substrate can also to absorb neutrons. Consider kink polarizer as part 1 (see Supplement II and [7]).

The potentials of the supermirror layers for the spin component inverse to the guiding field are close to each other when the magnetic layers are in the remanent state. At the same time, these potentials are close to the potential of the substrate material and do not exceed its value. Parts of the polarizer have reverse magnetization with respect to each other, and one of them is parallel to the guiding field (see Fig. 2). Neutrons of both spin components leave the 1st part of the polarizer. From the 2nd part only neutrons of the inverse spin component come out.

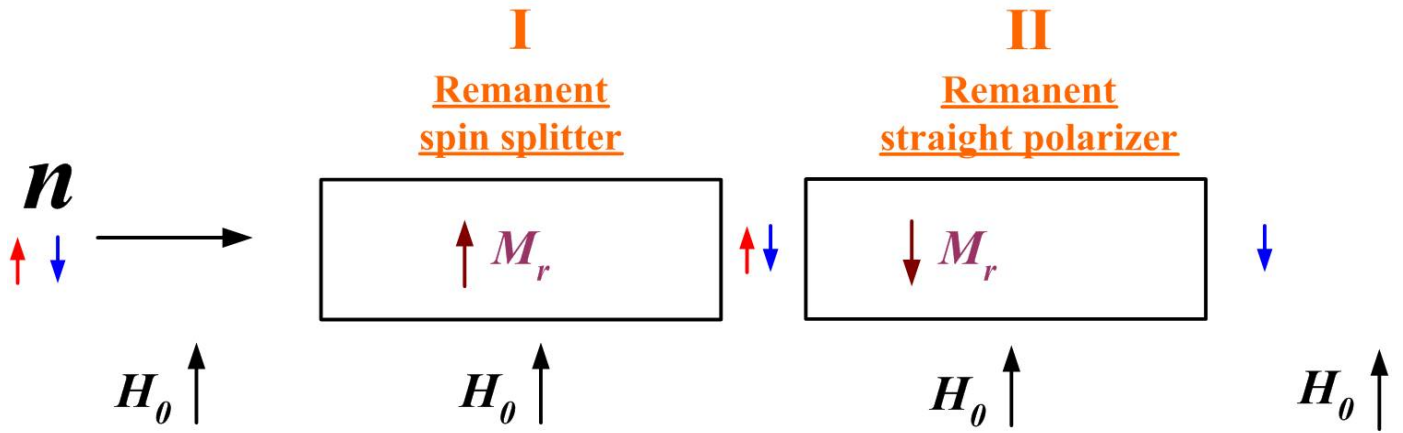

**Fig. 2.** Scheme of the new supermirror neutron polarizer.

The reflectivities curves $R^+$ and $R^-$ for both spin components of the beam for remanent polarizing supermirrors ($m = 2$) located in the 1st and 2nd parts of the polarizer are shown in (Fig. 3a) and (Fig. 3b), respectively.

As follows from the figures, the curves are inverse to each other.

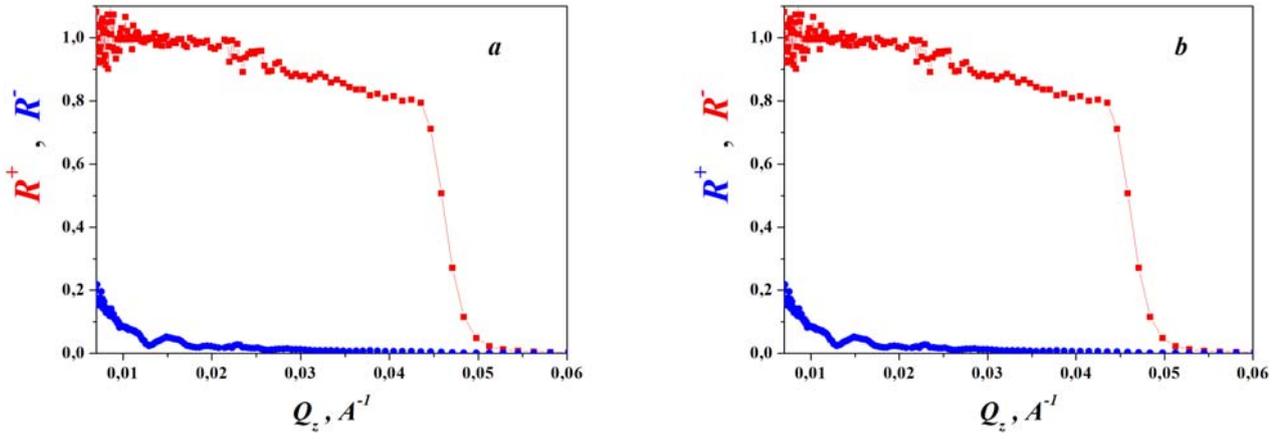

**Fig. 3a, b.** The reflectivities curves $R^+$ and $R^-$ for both spin components of the beam for remanent polarizing supermirrors ($m = 2$) located in the 1st (Fig. 3a) and 2nd (Fig.3b) parts of the polarizer.

Let us consider in detail the passage of each spin component of the beam through this polarizer using Figs. 4, 5. The beam profile at the input is the same for both spin components.

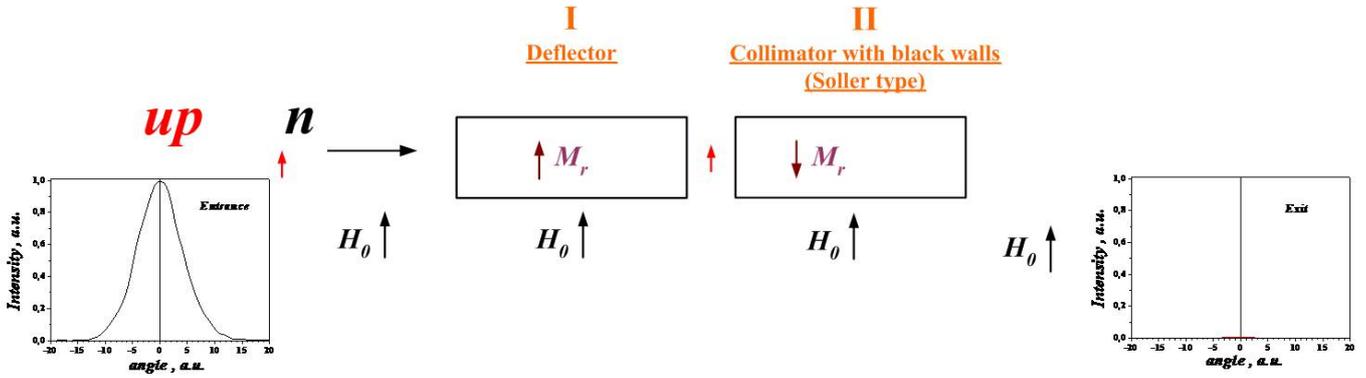

**Fig. 4.** Scheme of passage (+) spin component of the beam through the polarizer.

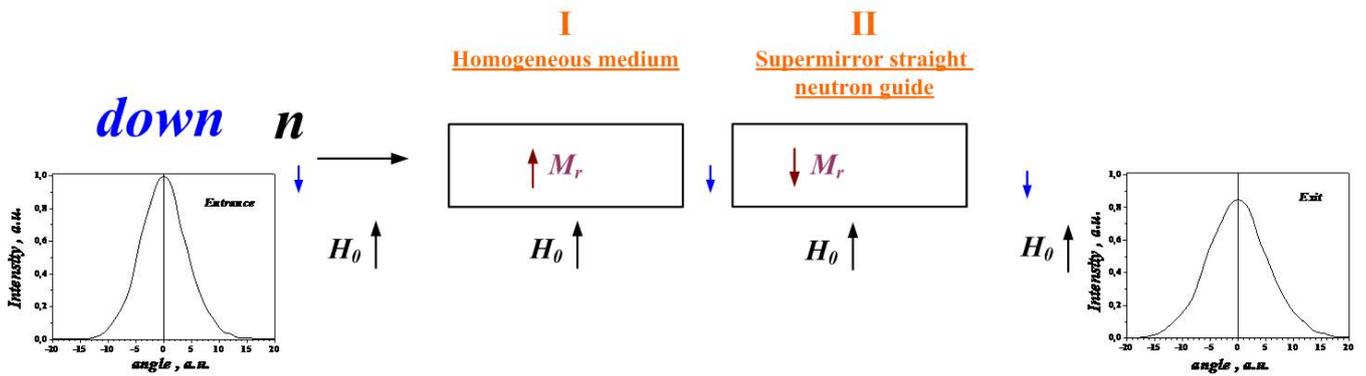

**Fig. 5.** Scheme of passage (-) spin component of the beam through the polarizer.

For (+) spin component of the beam (Fig. 4) the 1st part of the polarizer is a deflector, i.e. it significantly deflects the neutrons this component of the beam from their original trajectories due to the reflections of the neutrons of this spin component from the supermirror coatings in the channels of 1st part of the polarizer. In this case, the divergence of the beam increases significantly, and in the area of angles close to the beam axis falls to almost zero. When passing through the 2nd part of the polarizer, the neutrons of this spin component are reflected from the walls with a very low reflectivity, because the orientations of neutron spins and magnetization vectors of the magnetic layers of supermirrors are antiparallel.

The non-reflected neutrons of this spin component are absorbed in the sublayer of the supermirror. In fact, in this case, the 2nd part works as a conventional Soller collimator with walls made from neutron-absorbing and non-reflecting material. It would seem that neutrons must pass through it, because it has direct visibility in the angular range given by the width and length of each of its channels. But there are no neutrons in this angular range, because they were deflected from their original trajectories in the 1st part of the polarizer. As a result, there are practically no neutrons (+) spin component at the exit of the polarizer. Thus, the polarizer deflects in 1st part of it the neutrons of this spin component, and then absorbs them in the 2nd part. Thus, for this spin component of the beam, the polarizer performs two functions: **deflect and absorb!**

For neutrons (-) spin component of the beam, the situation is different (see Fig. 5). For them, the 1st part of the polarizer is a homogeneous medium, which they pass without deviations from their original trajectories, only slightly reducing their intensity due to absorption in the substrate material.

This is due to the fact, as mentioned earlier, that the neutron-optical potentials of the supermirror layers and the substrate material are close for spin component of the beam which inverse to the guiding field. The 2nd part of the polarizer, these neutrons pass through a conventional straight polarizing neutron guide reflecting from its walls with a high reflectivity (Fig. 3b), because in this case the spins of neutrons and the vector of magnetization of magnetic layers of the supermirror are parallel to each other. As is known, the beam does not change its divergence when passing through a straight neutron guide.

Thus, at the exit of the polarizer, the angular profile of the beam of this spin component repeats the beam profile at the entrance to the polarizer with a slight attenuation due to absorption in the substrate material (Fig. 5). As a result, the nonpolarized neutron beam passing through both parts of the polarizer will have a high negative polarization at the exit of the polarizer.

It is worth noting an important property of this polarizer. It can polarize beams having very wide angular distributions! The restriction is imposed by the parameter $m$ of polarizing supermirrors from the 2nd part of the polarizer and the level of their reflectivity. For example, $m = 5.5$. For neutrons with a wavelength of 5 A, the angular divergence of the beam can be $\pm$ 2.7 degrees. But, for supermirrors from the 1st part, the parameter $m$ can be small and equal to, for example, 2.

As the 1st part of this polarizer, in addition to kink polarizer, you can use polarized transmission bender without collimator [4]. Thus, there are two main variants of this solid state supermirror polarizer: 1) Ist part – kink polarizer, IInd part – polarizing straight NG, 2) Ist part – transmission polarizing bender, IInd part – polarizing straight NG.

Let's consider the first variant of implementation of the proposed polarizer, shown in Fig. 6. The polarizer consists of two parts: I-st part - kink polarizer (see Supplement II and [7]), II-nd part - straight polarizing neutron guide. The entire polarizer is in the magnetic guide field $H$. Both parts are in the remanent state: for the I-st part of the induction vector is oriented along the field, and the second part of the polarizer induction vector antiparallel to the vector of guiding field $H$. In Fig. 7 and 8 the schemes of one channel for the 1st and 2nd part of the polarizer are shown, respectively. Supermirror polarizing coating and parameter *m* of it can be the same for both parts (*Fe/Si*, *CoFe/TiZr*...).

For the 1st part, it is a coating without a neutron absorbing sublayer, and for the 2nd part of it must be with an absorber (*Gd*, *TiZrGd* ...). The material of substrate is transparent for neutrons - silicon, quartz, sapphire .... For the (-) spin component of the beam, the potentials of the supermirror layers must be close to each other in the remanent state and their value must not exceed the potential of the substrate material.

The neutrons with the spins parallel the field in Fig. 6 are shown with red arrows and against the field with black arrows. For clarity, the arrows are shown in the plane of the figure. In accordance with the reasoning presented above (see explanations to Fig. 4 and 5), neutrons along the field (red arrows) when passing through the I-st part will be reflected from the supermirror coatings of the channels and significantly deviate from their original trajectories including neutrons located near the axis of the angular distribution. Passing through the 2nd part, the neutrons of this spin component will be reflected from the walls of the straight polarizing neutron guide with a very low reflectivity ($R^-$).

As a result, only a small part of the neutrons of this spin component will come out of this neutron guide. Neutrons with spins against the field (black arrows) when passing through the I-st part practically will not be reflected from the supermirror coatings of the channels, because the potentials of the materials of the supermirror coating layers and the channel material are close to each other for this spin component. Therefore, these neutrons will not deviate from their original trajectories, i.e. the width of the angular profile of the transmitted beam will be the same as that of the beam at the input. Further, neutrons will pass through the straight polarizing neutron guide, both reflecting from its walls with a high reflectivity ($R^+$), and without reflections (straight-flight neutrons). At the exit from the 2nd part, the width of the angular profile of the neutron beam of this component will remain the same, since, as you know, when the beam passes through a straight neutron guide, its divergence does not change.

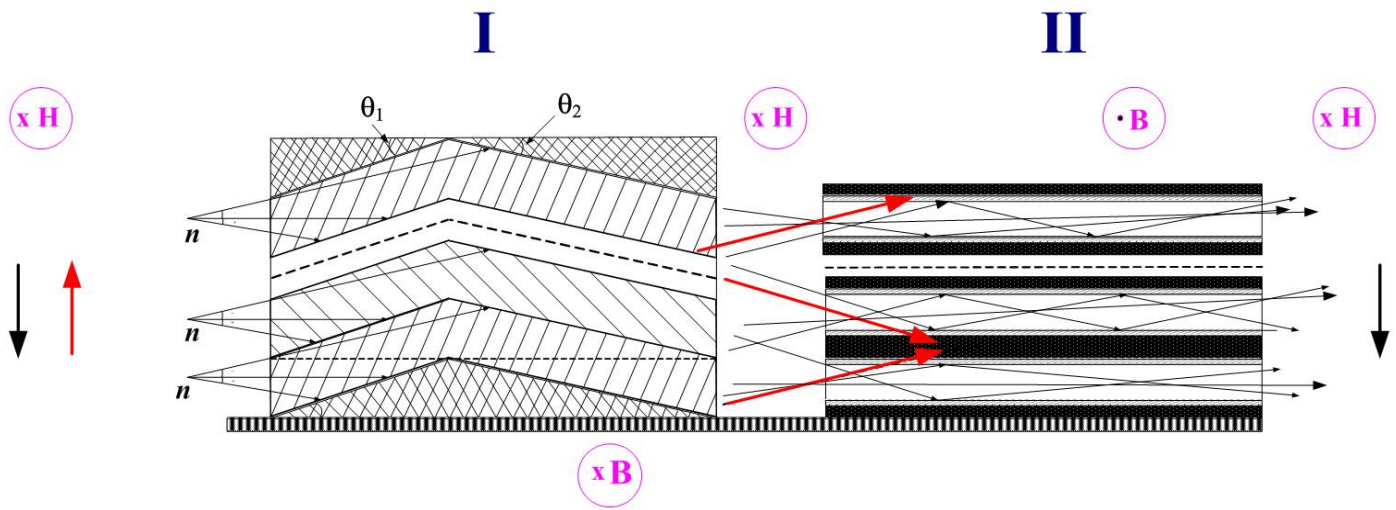

**Fig. 6.** New neutron solid state supermirror transmission remanent polarizer consisting from two parts: I – kink polarizer, II – straight polarizing neutron guide.

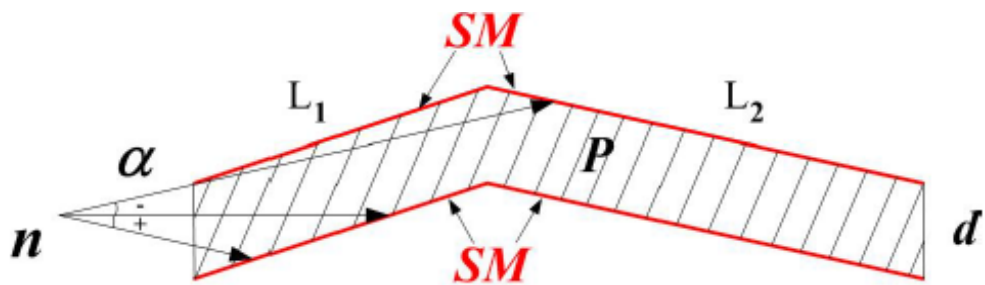

**Fig. 7.** The scheme of one channel for the 1st part of the polarizer - kink polarizer.

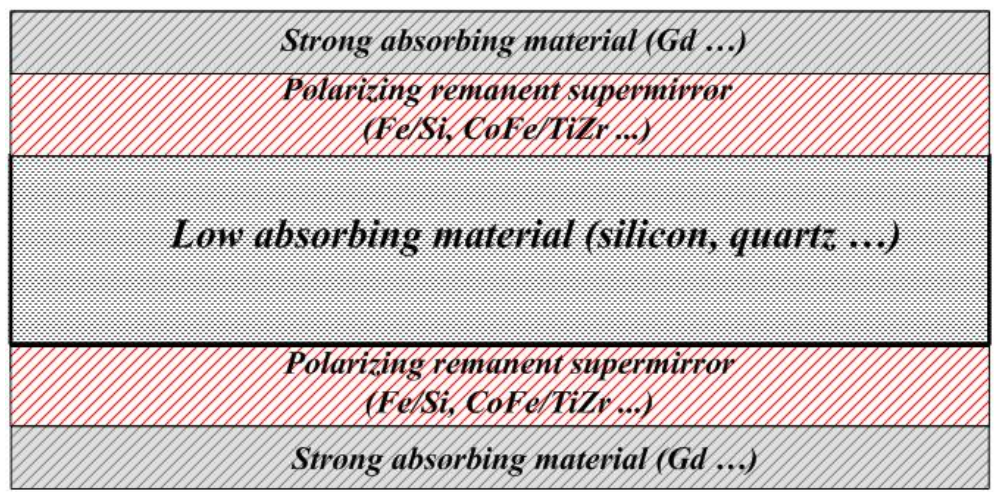

**Fig. 8.** The scheme of one channel for the 2-nd part of the polarizer – straight polarizing neutron guide.

Consider the first variant of the proposed polarizer (Fig. 6) (I-kink polarizer, II-straight polarizing NG) for the specific case of using remanent *CoFe/TiZr* ($m = 2$) supermirror coating and silicon as the material of channels for both parts of the polarizer. Geometric parameters of the channel of the 1st part of the polarizer: $d = 0.3$ *mm* - channel width or thickness of silicon plates, $\theta_1 = 15$ *mrad*, $\theta_2 = 10$ *mrad*, $L_1 = 20$ *mm*, $L_2 = 30$ *mm*.

In this part of the polarizer, as already noted, the absorbing *TiZrGd* sublayer was not used. Geometrical parameters of the channel of the 2nd part of the polarizer: $d = 0.3\ mm$ - channel width or thickness of silicon plates, $L = 60\ mm$ - channel length or length of silicon plates. The collimation angle (FWHM) of the 2nd part of the polarizer $\alpha$ is $\alpha = d/l = 5$ mrad, i.e. in this angle neutrons can pass without reflections from walls (straight-flight neutrons) through this collimator-polarizer. This part of the polarizer used an absorbing *TiZrGd* sublayer. The total length of the polarizer is 110 mm, the polarizing beam cross-section is 30*30 mm².

Using these geometric parameters, calculations have been carried out for a variant of a multichannel remanent transmission compact polarizer consisting of two parts: kink polarizer + straight polarizing neutron guide (collimator-polarizer) for a wavelength of *5.5 Å*.

On Fig. 9 the angular dependences of reflectivity curves for both spin components of the monochromatic beam with wavelength *5.5 Å* for *CoFe/TiZr* ($m = 2$) supermirror No. 86 (see Supplement II) are presented for both parts of polarizer and for remanent state in field 15 Oe.

The designations of the curves in Fig.9 are following $R^+$ and $R^-$ for the 1st part of the polarizer, and $R^+$ and $R^-$ for the 2nd part of it. In the 2nd part (straight polarizing neutron guide), the curves will be reversed for the spin components, because the magnetization there will be reversed.

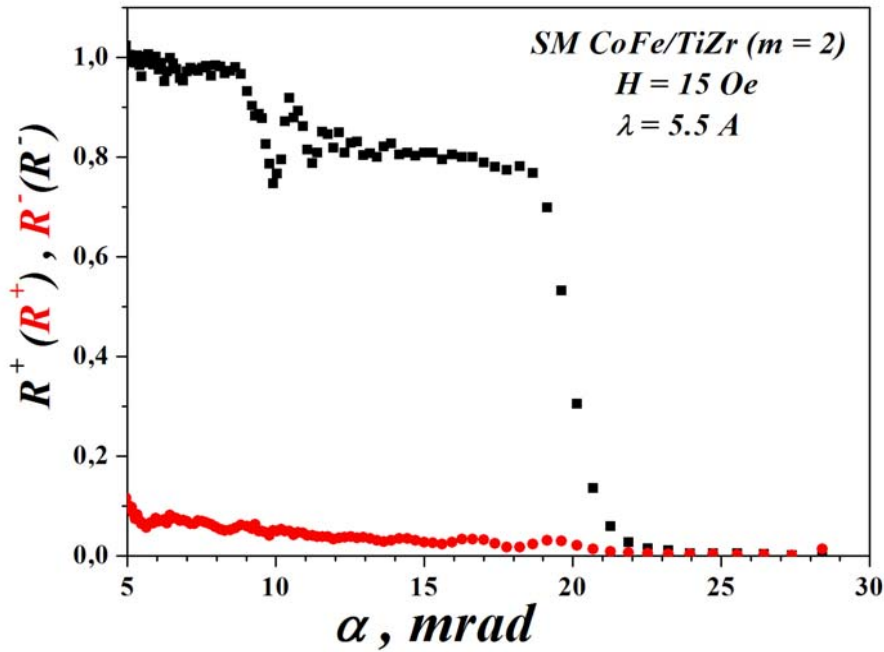

**Fig. 9.** The angular dependences reflectivity curves for both spin components of the monochromatic beam with wavelength 5.5 Å for *CoFe/TiZr* ($m = 2$) supermirror No. 86 (see Supplement II) for remanent state in the field 15 Oe. The designations of the curves in this figure are following $R^+$ and $R^-$ for the 1st part of the polarizer, and $R^+$ and $R^-$ for the 2nd part of it.

The scheme of one of the channels of the 2nd part – the straight polarizing neutron guide is shown in Fig. 10.

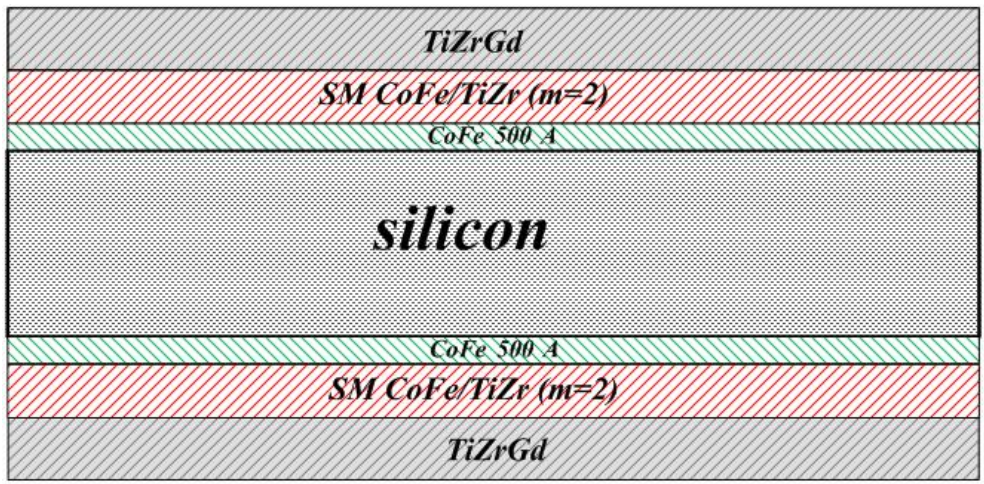

**Fig. 10.** The scheme of one of the channels of the 2nd part – the straight polarizing neutron guide.

In Fig. 11 calculated dependencies on angle $\alpha$ of normalized intensities $I^+$ and $I^-$ for, respectively, (+) and (-) spin components of monochromatic neutron beam with $\lambda = 5.5$ Å transmitted through the polarizer are shown for remanent state of *CoFe/TiZr* supermirror ($m$ = 2) No. 86 for field 15 Oe. Dependencies on $\alpha$ of intensity at the entrance of the polarizer $I_0$ and polarization $P$ of the transmitted beam are also shown in Fig. 11.

In calculations the curves of reflectivity (+) and (-) spin component of the beam from supermirror No. 86 in the remanent state were used for $H$ = 15 Oe (see Supplement II). The partial double reflection neutrons from the walls of 2nd part of the polarizer was taken account as well. As the picture shows, we have a beam only one spin component slightly weakened due to absorption in silicon. The width of the profile of this beam is almost the same as the width of the beam profile at the entrance. The beam profile of (+) spin component is significantly suppressed near the beam axis. Therefore, the beam polarization in this region is high and close to -1. Almost all peaks at ± 25 mrad (see Supplement II) disappeared due to multiple reflections from the walls of the collimator with reflectivity not exceeding 0.02 and the peak centered at $\alpha$ = 8 mad was largely suppressed. The average peak transmission of the beam (-) of the spin component $<T^->$ and the polarization of the exit beam $<P>$ are 0.58 and 0.99, respectively.

The transmittance of the polarizer will increase and its length will decrease if to use silicon wafers of smaller thickness. For example, with a thickness of 0.15 mm, the total length of the polarizer will be only 55 mm with the same polarizing beam cross-section and the same supermirror parameter $m$ = 2. On quartz, the absorption is still less than that of silicon. Although the result obtained is very good, the output beam parameters can be further improved (for example, to eliminate a small dip in the polarization at the angle of 8 mrad) by using supermirror No. 104 instead of supermirror No. 86 (see Supplement II), because at No. 104 the level of integral polarization **i**n the field of 20 Oe is 0.945 (against 0.9 at the supermirror No. 86).

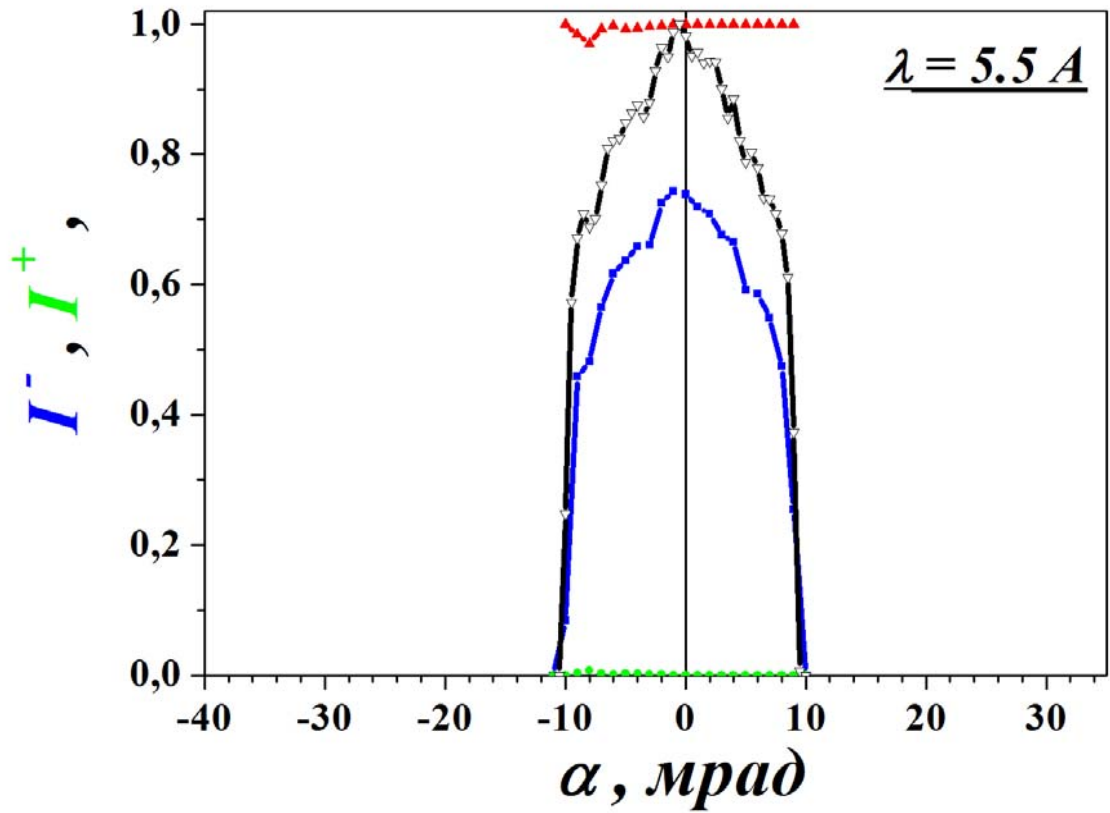

**Fig. 11.** Calculated dependencies on angle $\alpha$ of normalized intensities $I^+$ and $I^-$ for, respectively, (+) and (-) spin components of monochromatic neutron beam with $\lambda = 5.5\,\text{Å}$ transmitted through the polarizer for remanent state of *CoFe/TiZr* supermirror ($m = 2$) No. 86 in field 15 Oe. Dependencies on $\alpha$ are also shown of intensity at the entrance of the polarizer $I_0$ and polarization $P$ of the transmitted beam.

## 4. Operation of the proposed polarizer in saturating magnetic fields.

One of the variants for effective use of the proposed polarizer, consisting of two parts is the work of both its parts in saturating magnetic fields. The scheme of such a polarizer is shown in Fig. 12. As follows from the figure, the magnetization vectors of both parts of the polarizer are parallel to each other and parallel to the vector of magnetic guide field. A spin-flipper is installed between the parts of the polarizer. It is always in the state "on", i.e. it turns the spins of neutrons that have come out of the 1st part of the polarizer.

Thus, the action of this flipper is equivalent to the fact that the 2nd part of the polarizer, as it were, is in a state with reverse magnetization. Therefore, the properties of this variant of the polarizer and the principle of its operation are the same as for the remanent polarizer described above in the case of mutually reverse magnetization of both parts of the polarizer in the remanent state. The length of this variant of the polarizer will be slightly longer due to the use of a spin-flipper, but its main parameters will be slightly higher than for the remanent variant due to the magnetization to saturation of the magnetic layers of the supermirror structures of both parts of the polarizer.

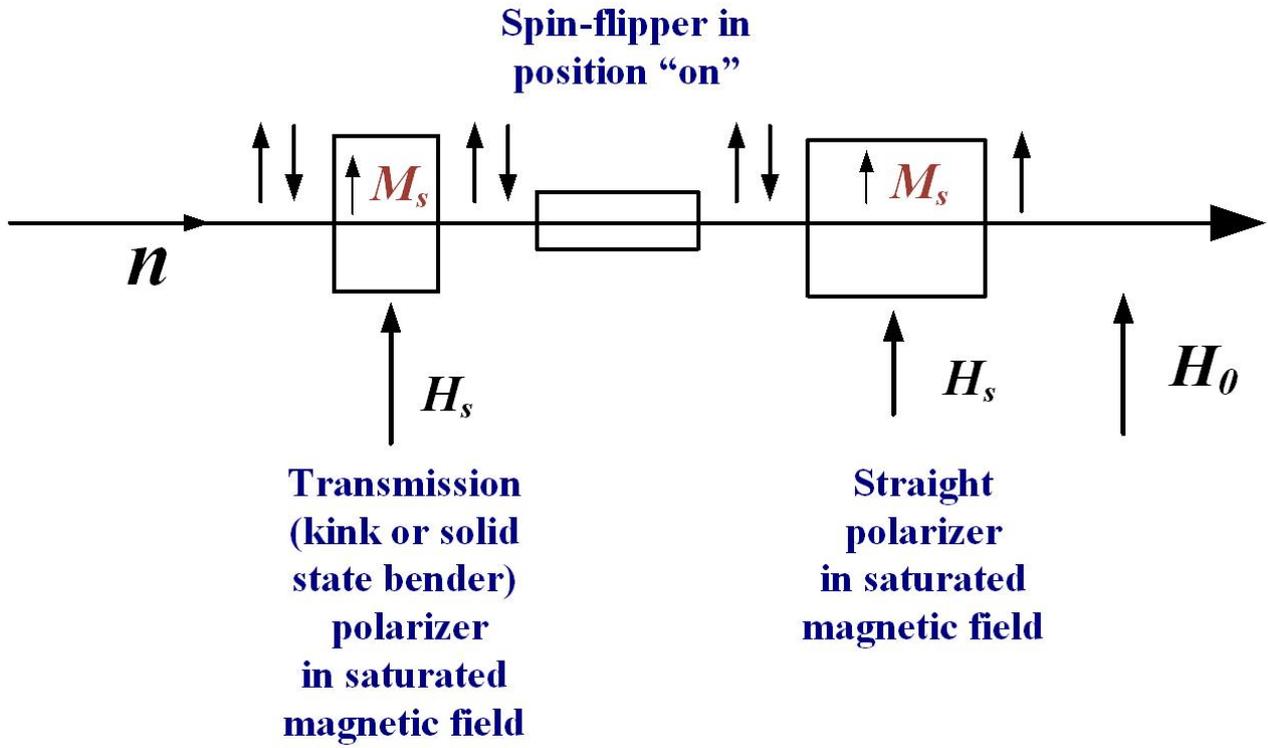

**Fig. 12.** Scheme of using the new polarizer in saturating fields.

In frames of scheme shown in Fig. 12, calculations were made for a wavelength of $\lambda = 5.5\,\text{Å}$ for a variant of a multichannel transmission compact polarizer consisting of two parts (kink polarizer + straight polarizing NG (collimator-polarizer)) located in saturating magnetic fields. The parameters of both parts are exactly the same as in the variant discussed above.

The angular dependences reflectivity curves for both spin components of the monochromatic beam with wavelength $\lambda = 5.5\,\text{Å}$ for *CoFe/TiZr* ($m = 2$) supermirror No. 86 (see Supplement II) are shown in Fig. 13 for saturating state in the field 470 Oe. The designations of the curves in this figure are following $R^+$ and $R^-$ for the 1st part of the polarizer, and $R^+$ and $R^-$ for the 2nd part of it.

In the calculations of the parameters of the polarizer presented in Fig. 12, the reflectivity curves shown in Fig. 13, were used.

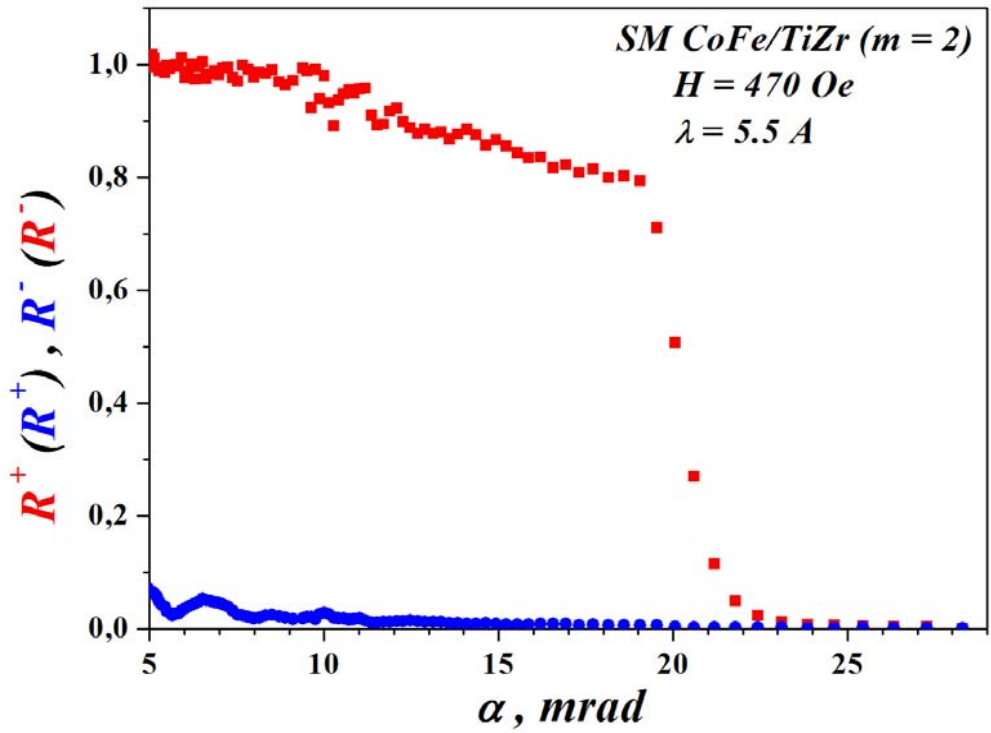

**Fig. 13.** The angular dependences reflectivity curves for both spin components of the monochromatic beam with wavelength $\lambda = 5.5\,\text{Å}$ for *CoFe/TiZr* (*m* = 2) supermirror No. 86 (see Supplement II) for saturating state in the field 470 Oe. The designations of the curves in this figure are following $R^+$ and $R^-$ for the 1st part of the polarizer, and $R^+$ and $R^-$ for the 2nd part of it.

Calculated dependencies on angle $\alpha$ of normalized intensities $I^+$ and $I^-$ for, respectively, (+) and (-) spin components of monochromatic neutron beam with $\lambda = 5.5\,\text{Å}$ transmitted through the polarizer are shown in Fig. 14 for saturated state of *CoFe/TiZr* supermirror (*m* = 2) in field 470 Oe. Dependencies on $\alpha$ of intensity at the entrance of the polarizer $I_0$ and polarization $P$ of the transmitted beam are also shown on this figure.

These curves are almost the same as those shown in Fig. 11, but the level of polarization of the transmitted beam is higher here than for the remanent variant. The average peak transmission for (-) spin component of the beam $<T>$ and the polarization of the beam on exit $<P>$ are 0.58 and 0.996, respectively.

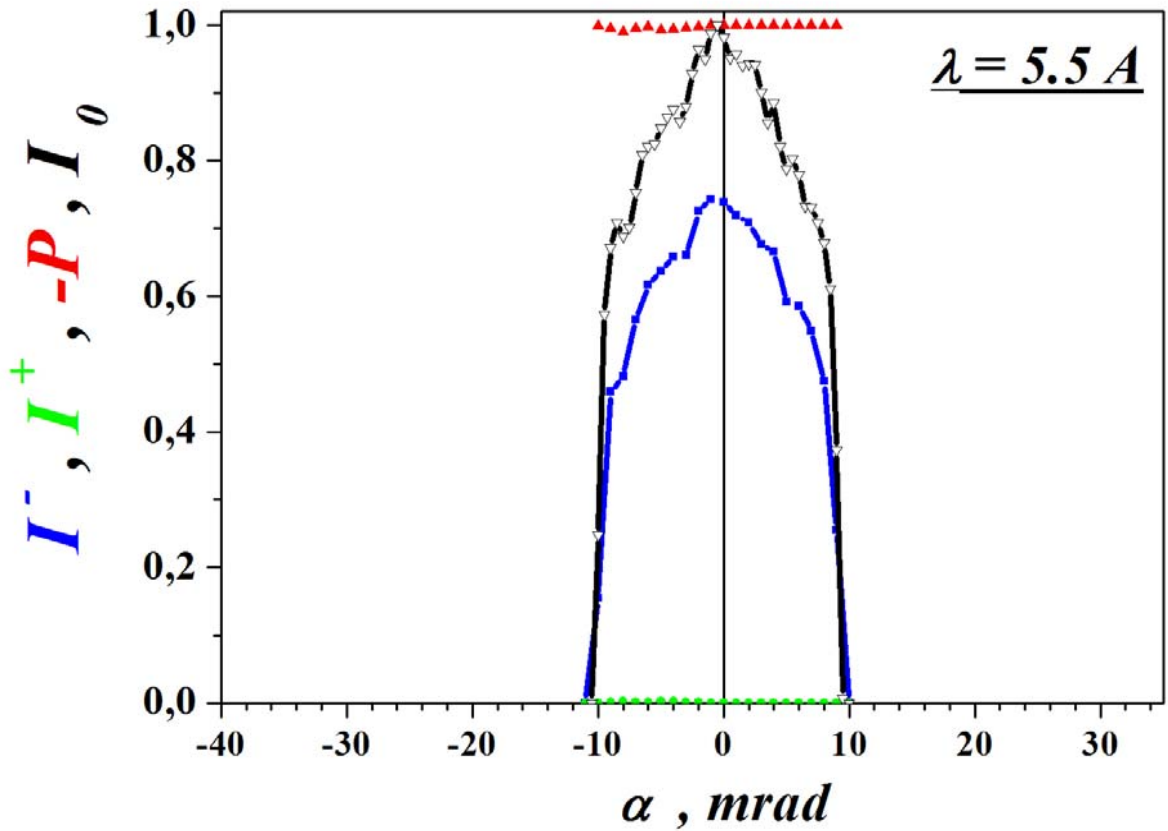

**Fig. 14.** Calculated dependencies on angle $\alpha$ of normalized intensities $I^+$ and $I^-$ for, respectively, (+) and (-) spin components of monochromatic neutron beam with $\lambda = 5.5$ Å transmitted through the polarizer for saturated state of *CoFe/TiZr* supermirror ($m = 2$) in field 470 Oe. Dependencies on $\alpha$ are also shown of intensity at the entrance of the polarizer $I_0$ and polarization $P$ of the transmitted beam.

## Conclusions:

I. A new solid state supermirror multichannel neutron polarizer is proposed.

II. Main features of the new polarizer:

II.1. It consists of two solid state polarizing parts having antiparallel magnetization of magnetic layers of supermirror structures. There are two possible configurations of the polarizer parts: 1) kink polarizer + straight polarizing NG, 2) transmission polarizing bender + straight polarizing NG.

II.2. Both parts of the polarizer have magnetic remanence. This allows it to be used in small magnetic fields.

II.3. It is a compact, because solid state.

II.4. The use of straight transparent plates for neutrons in the first configuration (kink polarizer + straight polarizing NG) parts of the polarizer allows it to be used as a polarizer and analyzer when working with large cross-section of the beams. Assembly of the device is greatly simplified, because it is not required to bend a large stack of short plates along the radius, but only to compress.

II.5. A wide angular range of the neutron beam is available for this polarizer.

II.6. The angular profile of the beam passing through the polarizer does not change!

II.7. The polarizer has a high transmittance for (-) spin component of the beam. The beam passed through the polarizer is highly polarized.

III. Examples of remanent *CoFe/TiZr* and *Fe/Si* supermirror coatings with $m = 2, 2.5$ are demonstrated in Supplement II. When using *Fe/Si* super mirrors with $m > 2.5$, additional research and development may need to be done to improve the remanent properties of polarizing supermirrors.

IV. A variant of effective use of the proposed polarizer with a spin-flipper for operation in saturating magnetic fields is proposed. This version of the polarizer has a slightly larger dimensions and here it is required to use saturating magnetic fields compared to the remanent variant, but the level of polarization of the transmitted beam is higher than for the remanent variant. In addition, this option can be used immediately without additional checks and studies.

## Acknowledgments.

The work was supported by the Ministry of Education and Science of the Russian Federation, Agreement No. 14.607.21.0194, unique identification number of project is RFMEFI60717X0194.

# Supplement I. Neutron transmission polarizers.

## I.1. V-cavity

The polarizer described in [1-3] is known. This is a V-cavity - neutron transmission supermirror polarizer. The main characteristic features of this polarizer are: the axis of the beam passing through the polarizer coincides with the axis of the incident beam and the high neutron transmittance for the main spin component of the beam. The scheme of one V-cavity channel and the principle of operation of this polarizer [3] is shown in Fig. I.1. Each of the channels of the V-cavity polarizer consists of a straight neutron guide and two identical long polarizing supermirrors on silicon substrates placed inside this neutron guide.

The walls of the neutron guide are coated with natural nickel ($m = 1$). Each of the mirrors consists of a set of rectangular polished silicon wafers, pressed against each other by the ends and lined up. The plates are coated with a polarizing (*CoFe/Si*, *Fe/Si*, etc.) supermirror coating ($m \geq 2$). These plates are oriented relative to each other so that they form two straight lines that close together, and the beam axis forms a small angle with each of the lines $\theta$, and the angle between the lines is equal $2\theta$. The angle $\theta$ is given by the ratio $\theta = \alpha_c \cdot \lambda_{min}$, where $\alpha_c$ is the critical angle of supermirror, $\lambda_{min}$ is the minimum wavelength in the neutron spectrum. The V-cavity is placed in a magnetic system that provides a saturating magnetic field. The set of plates is enclosed in a frame. Neutrons (+) spin components of the beam (i.e., the neutron spins, which are oriented in parallel to the vector of the magnetic field system, and the vector of magnetic guide field of facility) of the neutron falling on one of the plates of the V-cavity and are reflected from it at an angle less than critical. They are then reflected at an angle less than critical from the walls of the neutron guide with $m = 1$. As a result, the divergence of (+) spin component of the beam increases, which will lead to the absorption of these neutrons, or in the walls when falling on the wall at an angle greater critical, after reflection from the silicon mirror or in collimation system. Thus, in the beam passed through the V-cavity, the number of neutrons of (+) spin component will be significantly less than the neutrons (-) spin component, i.e. the passed beam will have negative polarization. In this case, the axis of the output beam coincides with the axis of the beam entering this polarizer. This is important because the transition from nonpolarized measurements to polarized neutron measurements does not require a laborious restructuring of the entire facility.

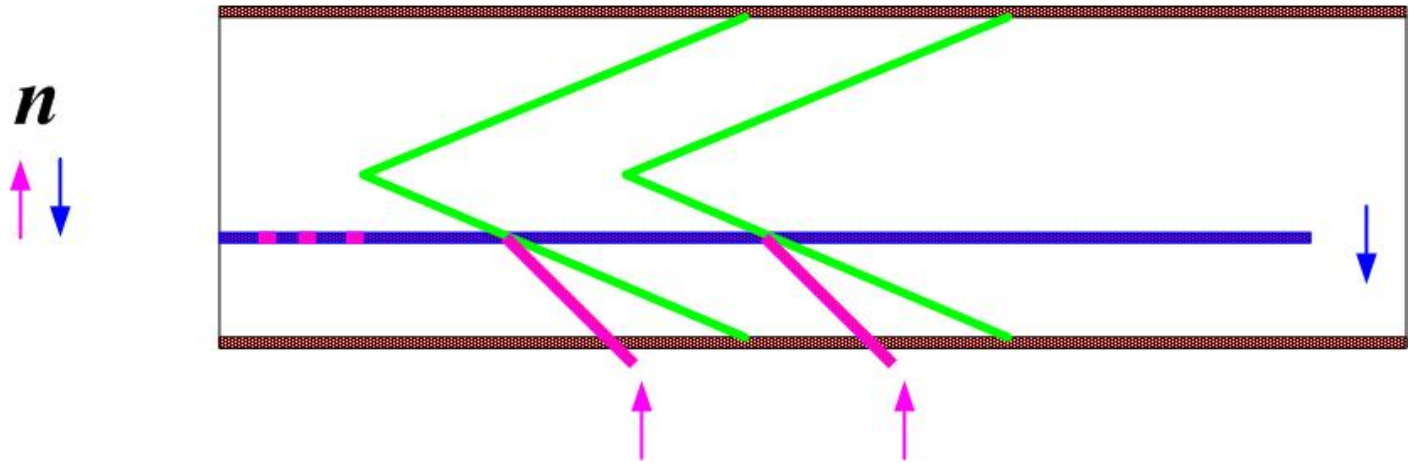

**Fig. I.1.** The scheme of one channel V-cavity and the principle of operation of this polarizer.

The disadvantages of the V-cavity polarizer are its long length ~ 0.5 – 0.6 *m* even when using polarizing supermirror coatings with high values of *m* ($m \sim 5$) and dips in the angular distribution of the beam intensity at the exit of the polarizer due to the separation glass plates between the polarizer channels.

**I.2. Solid state bender with collimator.**

From the literature, solid state supermirror neutron polarizers on neutron-transparent substrates including transmission polarizers are known.

On Fig. I.2 solid state neutron transmission polarizing bender with collimator [4] is presented. This polarizer is in a saturating magnetic field. The channels of bender and collimator are made from neutron-transparent silicon wafers. The polarizer plates are coated with a polarizing *CoFe/Si* supermirror coating, and the collimator plates are coated with absorbing layer of gadolinium. The neutron (+) spin component of the beam (purple arrows) passing through the bender deviates significantly from its original direction, as shown in the figure. When passing through the collimator installed at the bender outlet, this beam is completely absorbed in the non-reflective walls of the collimator. The neutron (-) spin component of the beam (blue arrows) passing through the bender does not deviate from its original direction, as shown in the figure, because the potentials of the supermirror layers and the wafer material are close. Therefore, this beam passes through the collimator and at its output the beam has a high negative polarization.

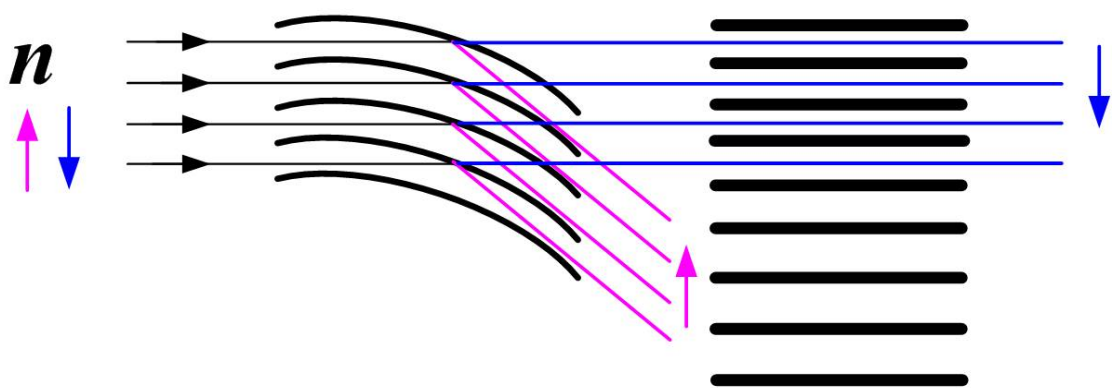

**Fig. I.2.** Solid state neutron polarizing transmission bender with collimator.

<u>Parameters of exit of this polarizer:</u> transmission for (-) spin state $T^- = 0.54$, polarization $P = -0.98$, angular range = 0.37 degree, wavelength = 4.72 Å. Parameters of bender: *CoFe/Si* (*m* = 2.3) supermirror on silicon wavers with thickness 0.16 mm, $L = 54$ mm, width = 20 mm. Parameters of collimator: silicon wavers (thickness 0.2 mm) with *Gd* coating, $L = 33$ mm.

<u>The advantages of this polarizer are:</u> the neutrons of the unwanted spin component of the beam are absorbed in the walls of the collimator, the axes of the input and output beams coincide.

<u>The disadvantages of this polarizer are:</u> small available angular range and low intensity of the exit beam.

**I. 3. Solid state neutron polarizing S-shaped benders.**

In paper [5] a supermirror solid state neutron polarizing S-bender was described. It is schematically shown in Fig. I.3. It is double-curved on the circumference of the multichannel neutron guide. The channels of this neutron guide are neutron-transparent silicon polished on both sides of the plate with a thickness of 0.15 mm. A polarizing *Fe/Si* (*m* = 3) supermirror coating is sputtered on both sides of each of these plates. Each coating is coated with a layer of gadolinium - neutron absorber. The results of the experiment show for the wavelength of 4.4 Å a homogeneous polarization of order $P = 0.987$ across the all cross-section of the bender and a high transmission 0.68 of (+) spin component of the neutron beam. The angular divergence of the beam at the output of this polarizer was 1.85 degrees. The length of this S-bender is 80 mm, the polarizing cross section is $30*100 mm^2$.

<u>Advantages of this polarizer:</u> does not deflect the axis of the beam passed through it, the unwanted spin component of the beam is absorbed in gadolinium layers.

<u>Disadvantages of solid state S-shaped bender:</u> the beam passed through it increases its divergence, difficulties in the manufacture a large cross-section of such polarizer.

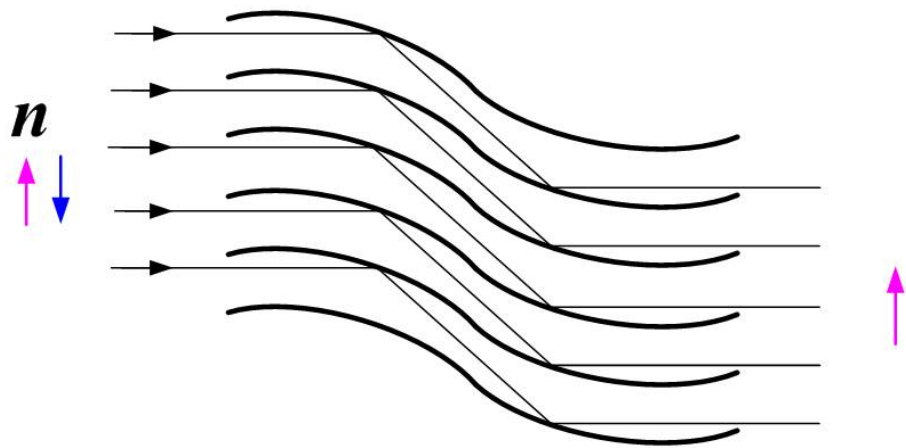

**Fig. I.3.** Scheme of solid state neutron polarizing S-shaped bender.

## I.4. Solid state compact neutron transmission kink polarizer

A new compact neutron supermirror transmission polarizer is proposed in [6, 7]. This polarizer considerably more compact then V-cavity! The polarizer consists of a set of plates transparent to neutrons placed in the magnet gap. There are no air gaps between the plates. Polarizing supermirror coating without absorbing underlayer is deposited on the polished surfaces of the plates. Magnetic and nonmagnetic layers of the supermirror coating as well as the material of the plates have nearly equal neutron-optical potentials for spin-down neutrons. There is a considerable difference between neutron-optical potentials of layers in the supermirror structure for spin-up neutrons. As a result, spin-up neutrons reflect from the supermirror coating and deviate from their initial trajectories whereas spin-down neutrons do not practically reflect from the coating and, consequently, do not deviate from their initial trajectories. Thus, spin-down neutrons dominate near the axis of distribution of intensity on the angle for the beam transmitted through this polarizer, i.e., the beam is substantially polarized.

The polarizer design (side view) is shown in Fig. I.4. A set of neutron transparent plates P is sandwiched between two broken polished metal surfaces 1 and 2 (the punch and the matrix) so that each plate of thickness $d$ forms a broken asymmetric neutron guide channel with supermirror walls. This channel is shown in Fig. I.5. As is shown in the figure, the channel height (plate thickness) is $d$, the front part of the channel (plate) has length $L_1$ and it makes with the beam axis angle $\theta_1$, and the corresponding parameters the output part of the channel are $L_2$ and $\theta_2$, respectively. The width of the channel along the axis perpendicular to the plane of the figure is equal to $L_3$ (not shown in the figure) is determined by the width of the incident beam.

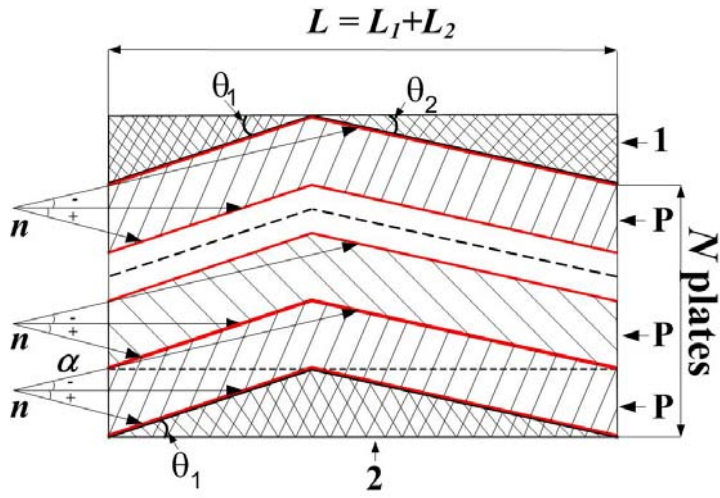 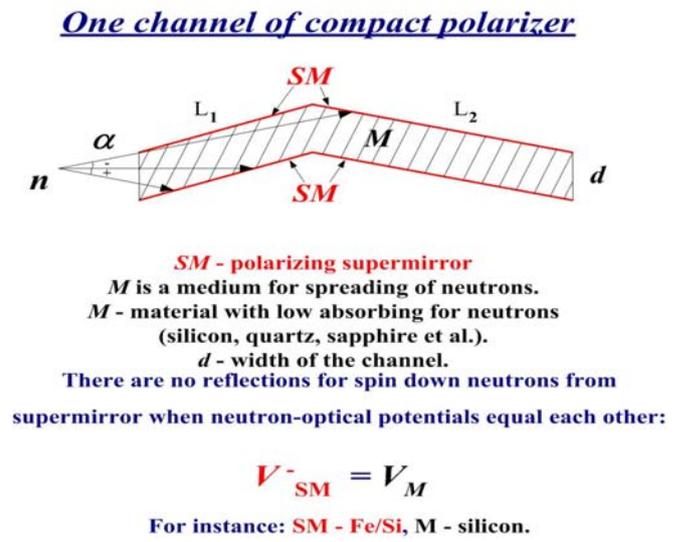

**Fig. I.4.**                                                   **Fig. I.5.**

**Fig. I.4.** General view of the proposed compact neutron transmission supermirror polarizer.
**Fig. I.5.** General view of one channel of the proposed compact transmission neutron supermirror polarizer.

      A divergent neutron beam **n** comes to the entrance of each channel (Figs. I.4 and I.5). Entrance angle $\alpha$ of neutrons is counted off from horizontal line. A polarizing supermirror coating < SM > covers all polished surfaces of width **d**. For neutrons with (-) spin component of the beam, the neutron-optical potentials are close to each other of the supermirror coating layers and the material of the plate so that the critical angle $\alpha_c^-$ is close to zero for the boundary "material-supermirror". Slight exceeding is allowed of the plate potential over potentials of layers of the supermirror. For neutrons with (+) spin component of the beam, the neutron-optical potentials of its layers differ significantly from each other so that the corresponding critical angle $\alpha_c^+$ is great for the same boundary. Polarizer plates are in the field of the magnetic system of the polarizer. This field lies in supermirror plane and perpendicular to the plane of the figure (Fig. I.5). The applied field saturates the magnetic layers of supermirrors deposited on the plates. The number of broken asymmetric channels (plates) and their width $L_3$ are defined by the required beam cross section used in the facility which is equal to $h \times L_3$, where $h$ is the beam height. $N$ such channels are used which are pressed to each other without air spaces, where $N$ is defined as $h = N \cdot d$.

      Results of calculations are shown in Fig. I.6 for $I^+$ and $I^-$ at $\lambda$ = 5.5 Å for the first polarizer dependence on $\alpha$, of intensity $I$ at the entrance of the polarizer and polarization $P = (I^+ - I^-)/(I^+ + I^-)$ of the transmitted beam. Neutron absorption in silicon has been taken into account. As expected, $P$ turns out to be negative and the beam divergence of (+) spin component is increased due to the reflection from supermirror walls of the polarizer. The latter circumstance is in the origin of complicated structure of angle distribution of $I^+$ which has three separated peaks. As it is seen from the figure, the area $\alpha \approx -6 \div 16$ *mrad* has small amount of neutrons of (+) spin component. In contrast, the width of the angle distribution of (-) spin component does not practically change upon coming through the polarizer. As a consequence, the

polarization of the transmitted beam in the angle range $\alpha \approx -6 \div 16$ mrad is extremely high being close to -1.

Calculated wavelength dependencies are shown in Figs. I.7 of polarization $P$ and transmission coefficient $T^-$ (in which the absorption in silicon is taken into account) of neutrons of (-) spin component for the polarizer. As it is seen from the Fig. I.7, $T^-$ is large and it decreases from 0.9 to 0.8 upon wavelength decreasing from $\lambda = 4.5 Å$ to $\lambda = 9.5 Å$. Curves for $P$ are obtained for angular divergences of the beam at the exit of polarizer of 8 mrad ($\pm 4$ mrad) and 10 mrad ($\pm 5$ mrad). As it is seen from these dependencies, the polarization is better than -0.99 in the whole spectral range $\lambda = 4.5 \div 10 Å$ and for $\lambda = 5.5 \div 10 Å$ for the divergence of 8 mrad and 10 mrad, respectively.

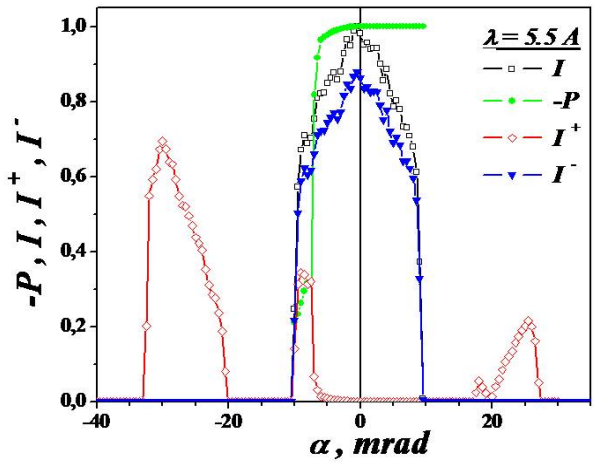 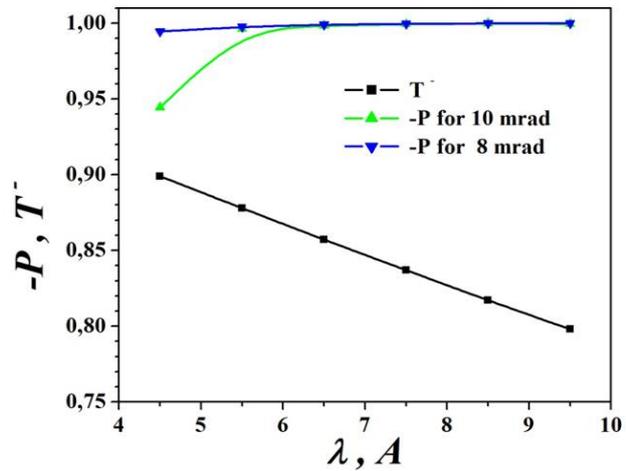

**Fig. I.6.**                                                             **Fig. I.7.**

**Fig. I.6.** Calculated dependencies on angle $\alpha$ of normalized intensities $I^+$ and $I^-$ for, respectively, (+) and (-) spin components of monochromatic neutron beam with $\lambda = 5.5 Å$ transmitted through the polarizer for saturated state of *CoFe/TiZr* supermirror ($m = 2$). Dependencies on $\alpha$ are also shown of intensity at the entrance of the polarizer $I_0$ and polarization $P$ of the transmitted beam.

**Fig. I.7.** Calculated wavelength dependencies of polarization $P$ and transmission coefficient $T^-$ (in which the absorption in silicon is taken into account) of neutrons of (-) spin component for the beam transmitted through the polarizer. Curves for $P$ are obtained for angular divergences of the beam at the exit of polarizer of 8 mrad ($\pm 4$ mrad) and 10 mrad ($\pm 5$ mrad).

Advantages of kink polarizer: the axis does not deflect the past through it of the beam, a short length of polarizer even at small values of the parameter *m* of supermirror coating, the possibility of polarized beams of large cross-sections, since do not need to bend the channels of the polarizer, very high polarization of the transmitted beam in the angular range of 8 mrad ($\pm 4$ mrad).

Disadvantages of solid state kink polarizer: insufficiently wide angular range of the beam on exit of the polarizer, in which the polarization is close to 1: as follows from Fig. I.6, in the angular distribution (+) of the spin component of the beam, 3 peaks are visible with centers at $\alpha = -30$, -8 and +25 mrad.

Recently, a compact polarizer on silicon plates with *CoFe/TiZr* ($m = 2$) supermirror coating was created at PNPI. 3D drawing of the polarizer is shown in Fig. I.8a. The photo of the polarizer is shown in Fig. I.8b. Its dimensions are small and equal to 45x55x75 $mm^3$. The polarizing cross section of the beam is enough 30x30 $mm^2$.

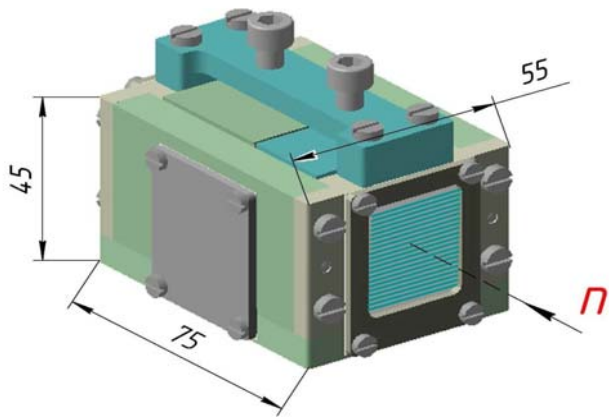
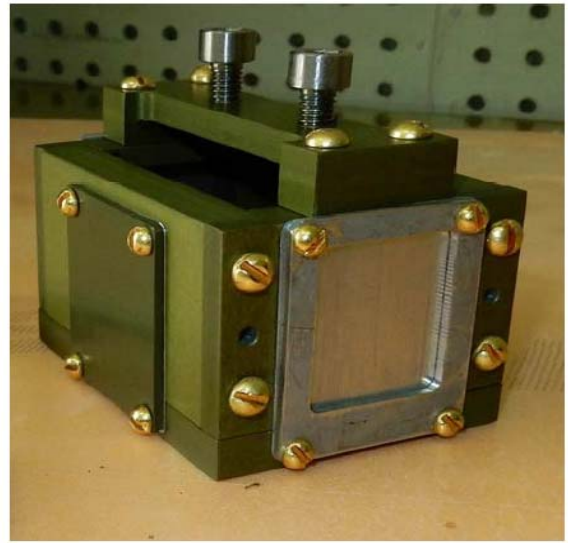

**Fig. I.8a.**  **Fig. I.8b.**

**Fig. I.8 a, b.** (a) 3D drawing of the polarizer, (b) the photo of the polarizer.

The results of measurements of this polarizer are presented at international conference ECNS 2019 [13] and will be present at international conference AOCNS 2019.

# Supplement II. Remanent neutron polarizing supermirrors.

## II.1. Remanent polarizing *CoFe/TiZr* supermirror of PNPI ($m = 2$ и $m = 2.5$).

The neutron polarizing *CoFe/TiZr* supermirrors on a glass substrate were prepared in Petersburg Nuclear Physics Institute (PNPI) [8] by magnetron sputtering. In order to obtain high polarization of the neutron beam reflected from such supermirror compositions of *CoFe* and *TiZr* alloys were chosen so that for neutrons (-) spin components of a beam of neutron-optical potentials of these alloys were close to zero. In order to minimize the reflection (-) of the spin component of the beam from the glass, an antireflective absorbing layer of the *TiZrGd* alloy was sputtered directly onto the glass substrate.

The curves of the dependence of the integral polarizing efficiency of $P_{int}$ on the magnetic field $H$ applied to this supermirror in the easy (1) and hard (2) directions are presented on Fig. II.1. Measurements were carried out on the neutron reflectometer NR-4M [14] on the "white" beam of thermal neutrons (maximum spectrum at a wavelength of 1.25 Å) at a glancing angle of 15 minutes. As can be seen from Fig. II.1, this supermirror has a remanent magnetization and it can be used in a field of the order of 25 Oe, because in this field the value of $P_{int}$ close to the similar value at a saturating field of 470 Oe.

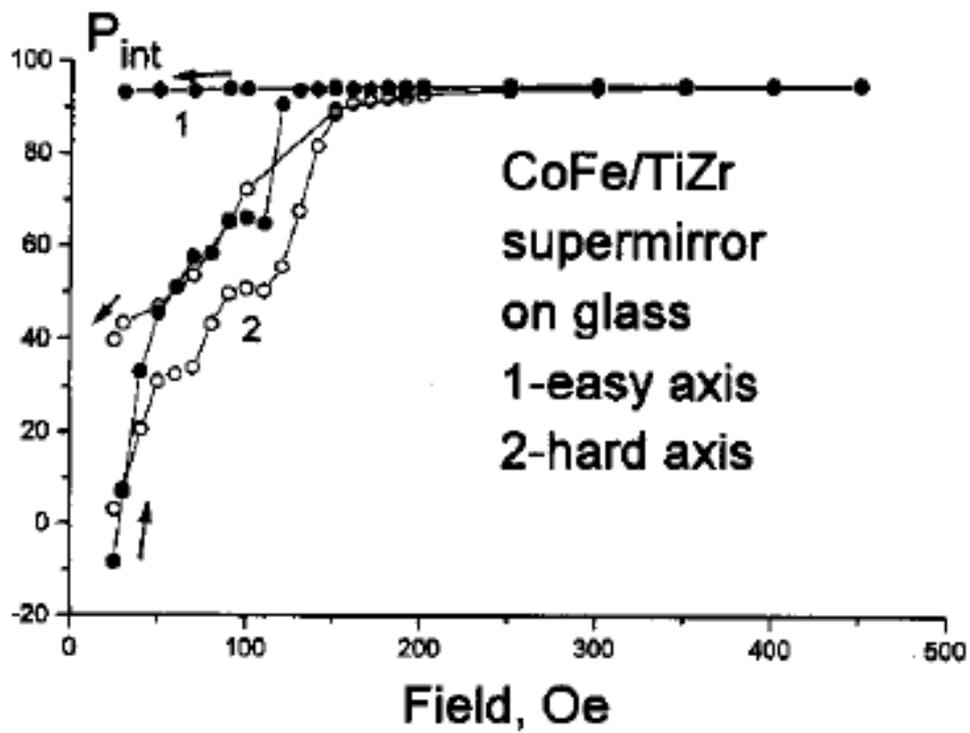

**Fig. II.1.** The polarizing efficiency of the $P_{int}$ CoFe/TiZr supermirror, measured on a "white" beam of thermal neutrons (maximum spectrum at a wavelength of 1.25 Å) on the reflectometer NR-4M of PNPI, as a function of the magnetic field applied to the supermirror $H$ in the easy (1) and hard (2) directions [8].

The curves of the dependence of the integral polarizing efficiency of $P_{int}$ on magnetic field $H$ applied to the supermirror No. 86 in the easy direction are presented in Fig. II.2. The curves were obtained in the same geometry at NR-4M. The level $<P_{int}>$ equals $+/-0.88$ for hysteresis curve in remanence region at $H = 15$ Oe.

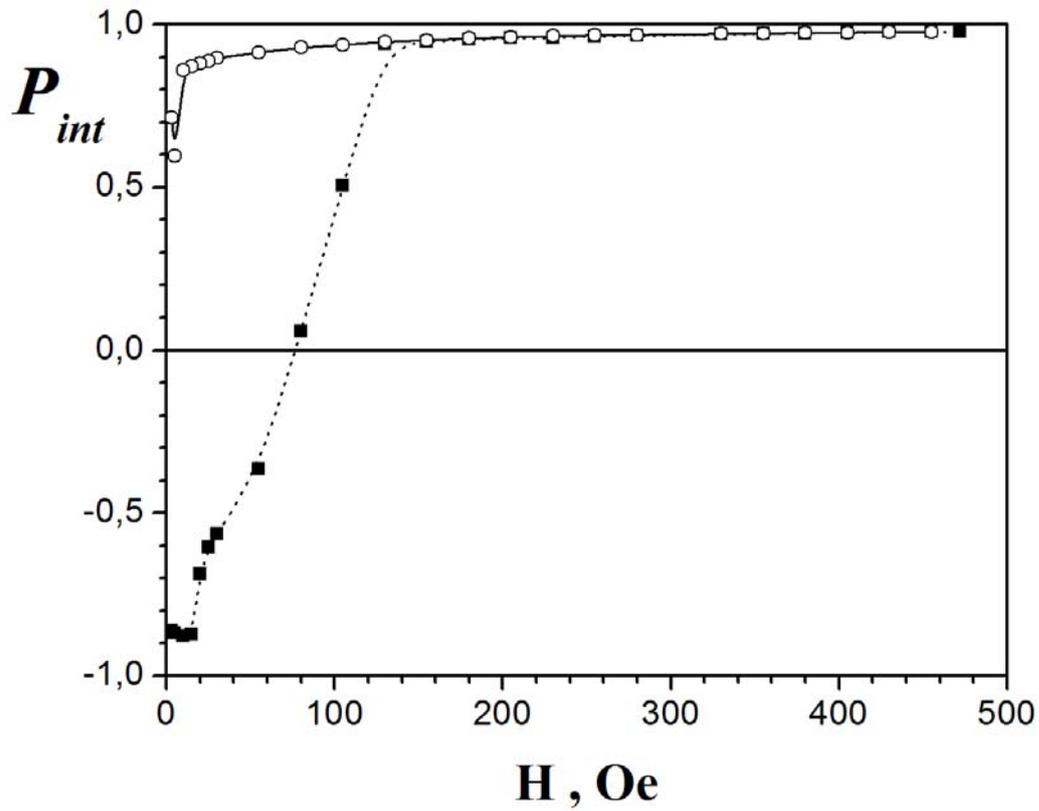

**Fig. II.2.** The polarizing efficiency of $P_{int}$ of *CoFe/TiZr* supermirror No. 86, measured on a" white " beam of thermal neutrons on the reflectometer NR-4M PNPI, as a function of the magnetic field $H$ applied to the supermirror in the easy direction.

The reflectivity curves $R^+$ and $R^-$ for both spin components of the beam for *CoFe/TiZr* supermirror No. 86 are presented in Fig. II.3 [11] for two magnitudes of magnetic field $H = 15$ and 470 Oe applied to this supermirror in the easy direction. Measurements were carried out by the time-of-flight method also on the neutron reflectometer NR-4M on the "white" beam of thermal neutrons.

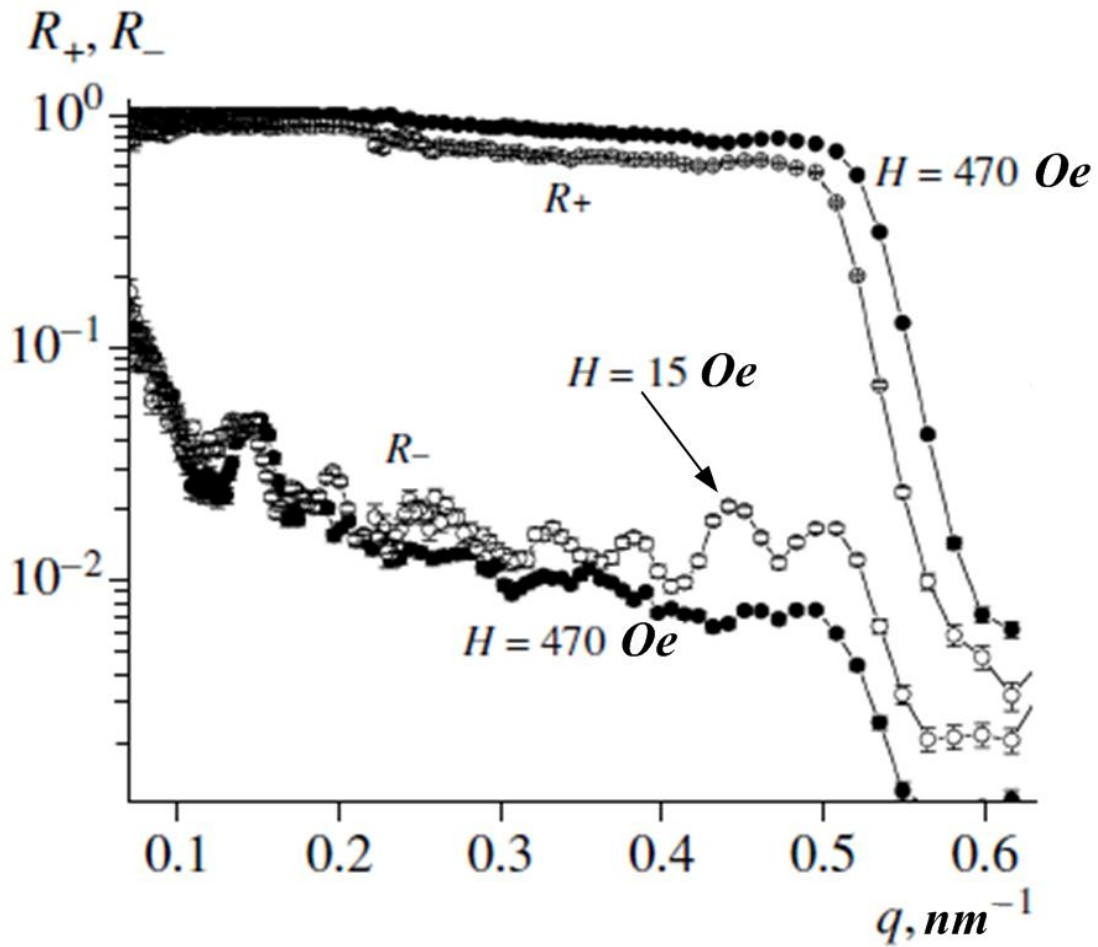

**Fig. II.3.** Spectral dependences of the reflection coefficients ($R^+$ and $R^-$) of both spin beam components on the *CoFe/TiZr* supermirror No. 86 for two magnitudes of the magnetic field $H$ = 15 and 470 Oe applied to a supermirror in the easy direction.

The curves of the dependence of the integral polarizing efficiency of $P_{int}$ on the magnetic field $H$ applied to the supermirror ($m$ = 2.5) No. 17 in the easy direction obtained in the same geometry at NR-4M are presented in Fig. II.4. The level $<P_{int}>$ for this supermirror is $\pm 0.9$ in the remanence region at $H$ = 30 Oe.

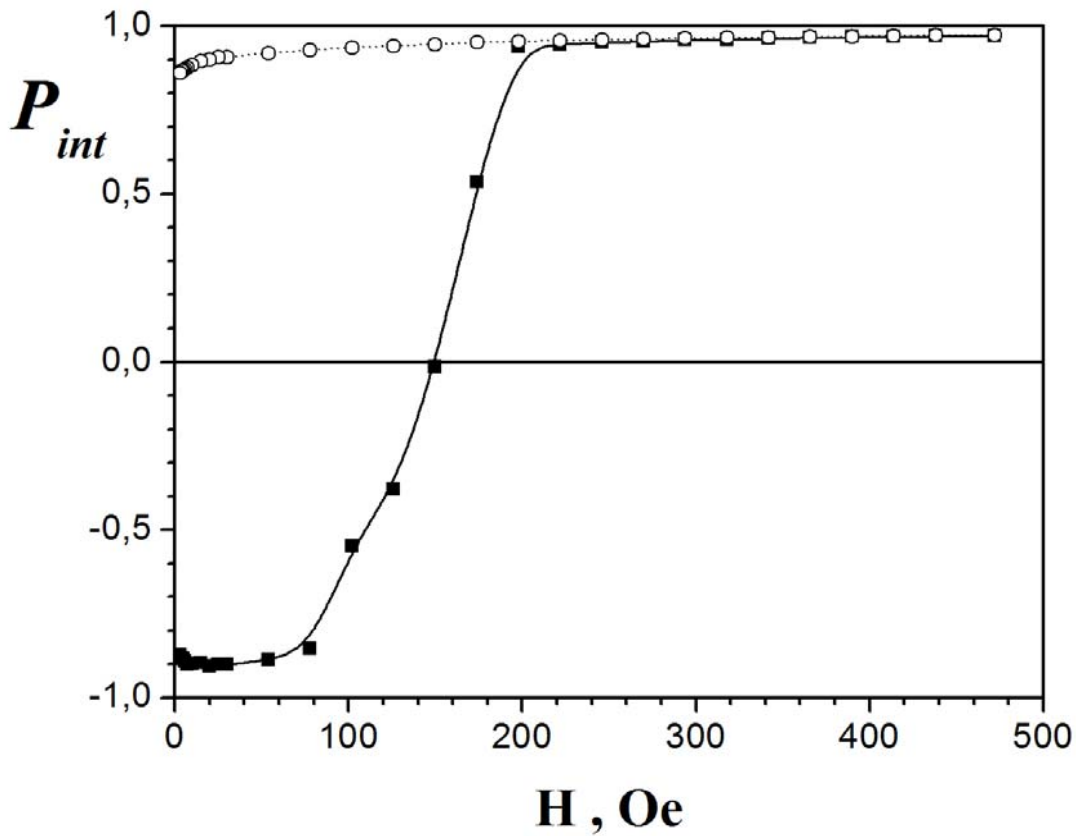

**Fig. II.4.** The polarizing efficiency of $P_{int}$ of *CoFe/TiZr* (*m* = 2.5) supermirror No. 17, measured on a" white " beam of thermal neutrons on the reflectometer NR-4M PNPI, as a function of the magnetic field *H* applied to the supermirror in the easy direction.

The curves of the dependence of the integral polarizing efficiency of $P_{int}$ on the applied field *H* in the easy direction to the *CoFe/TiZr* (*m* = 2) supermirror No. 104, obtained later at the same magnetron sputtering facility, are presented in Fig. II.5. Measurements were also carried out on the neutron reflectometer NR-4M on a "white" beam of thermal neutrons in the same geometry. As can be seen from Fig. II.5, this supermirror has a pronounced remanent magnetization and can be used in a field of about 20 Oe. So for the upper branch of the hysteresis loop at 20 Oe $P_{int}$ = 0.945, and at the same field for the lower branch we get $P_{int}$ = -0.945, whereas at the saturating value of the field in 470 Oe polarization is maximum and equal to 0.971. Thus, for a field of 20 Oe, the polarization difference from its maximum value is quite insignificant and it can be used in a small field of the order of 20 Oe. This mirror is much better than all the above considered.

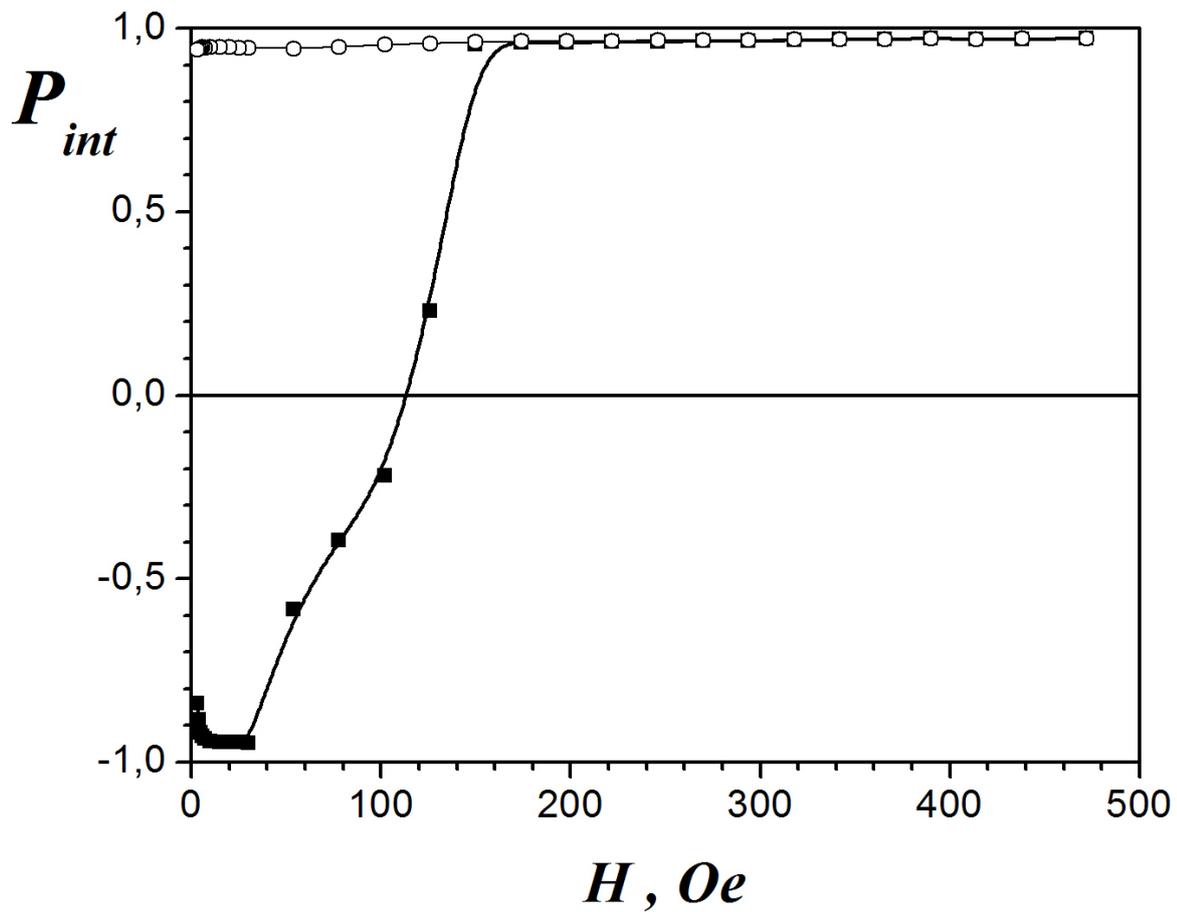

**Fig. II.5.** The polarizing efficiency of $P_{int}$ CoFe/TiZr ($m = 2$) supermirror No. 104, measured on a" white " beam of thermal neutrons on the reflectometer NR-4M PNPI, as a function of the magnetic field $H$ applied to the supermirror in the easy direction.

Spectral dependences of the reflection coefficients ($R^+$ and $R^-$) of both spin beam components on the *CoFe/TiZr* (*m* = 2) supermirror No. 104 are presented in Fig. II.6 for saturating magnitude of the magnetic field $H$ = 470 Oe applied to the supermirror in easy direction.

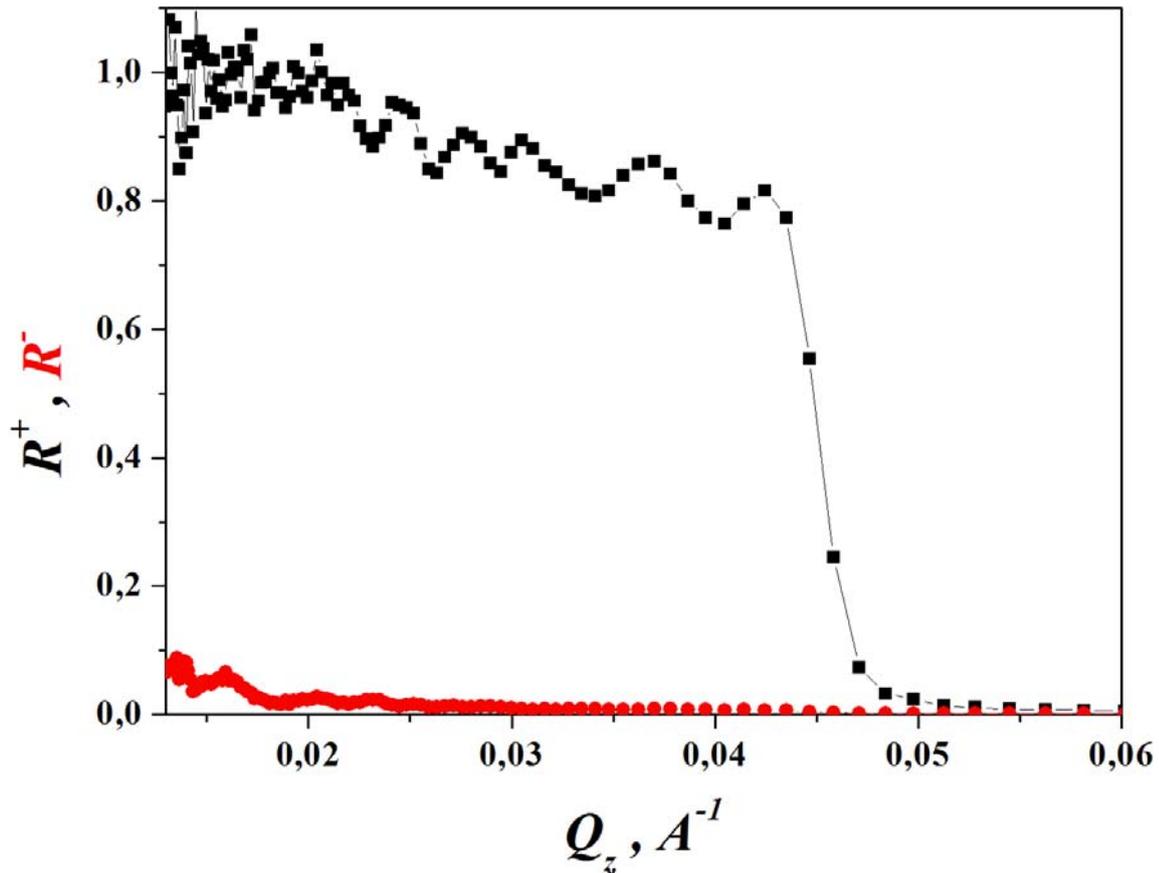

**Fig. II.6.** Spectral dependences of the reflection coefficients ($R^+$ and $R^-$) of both spin beam components on the *CoFe/TiZr* (*m* = 2) supermirror No. 104 for saturating magnitude of the magnetic field $H$ = 470 Oe applied to the supermirror in easy direction.

## II.2. Remanent neutron polarizing *Fe/Si* (*m* = 2) supermirror.

In paper [10] the results of investigations of remanent multilayer *Fe/Si* structures are presented. Hysteresis curves along the hard (1) and easy (2) magnetization axes for a multilayer periodic structure of 20Fe(70 Å)/Si(70 Å) are shown [10] in Fig. II.7. The easy magnetization axis for this structure corresponds to the perpendicular orientation of the substrate relative to the direction of motion during the magnetron sputtering process of the multilayer structure. As follows from the figure, the hysteresis curve for the easy magnetization axis has a pronounced rectangular shape, which indicates a high remanence of this structure. So for the field value $H \sim 20$ Oe, the magnetization values for the lower and upper branches of the loop are close to the maximum values of magnetization, but with different signs.

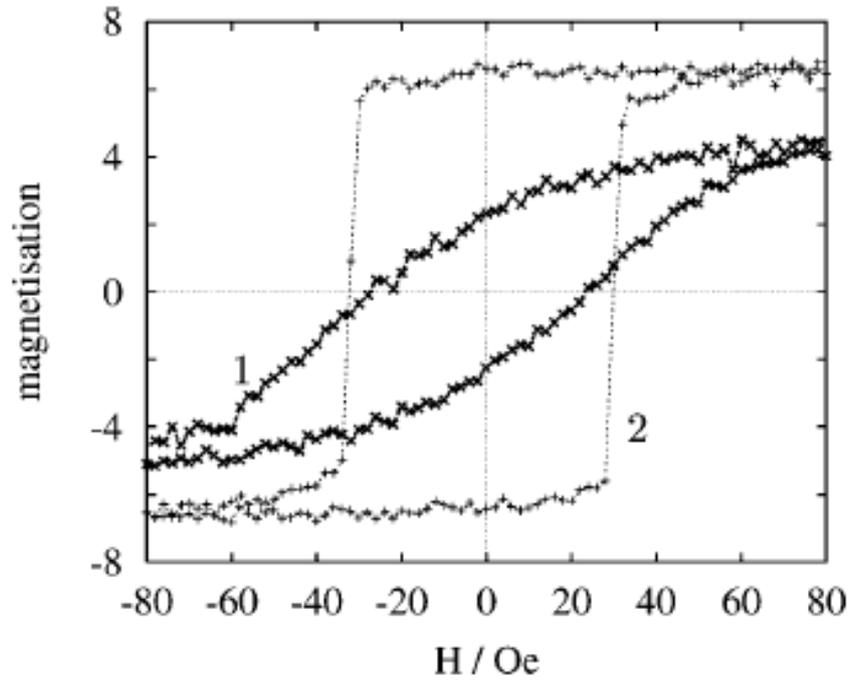

**Fig. II.7.** Hysteresis curves along the hard (1) and easy (2) magnetization axes for a *20Fe(70 Å)/Si(70 Å)* multilayer periodic structure [10].

The reflectivity and transmission curves for both spin beam components are shown in Fig. II.8, as well as the polarization curves for the *Fe/Si* supermirror ($m = 2$) in a small magnetic field ($H < 20$ Oe) after magnetization in a large field (800 Oe) [10]. As can be seen from the graphs, the polarization levels of the transmitted and reflected beams are high. So, for reflectivity curve, the polarization is at $P \sim 0.95$ in the range of angles from 0.6 to 0.8 degrees.

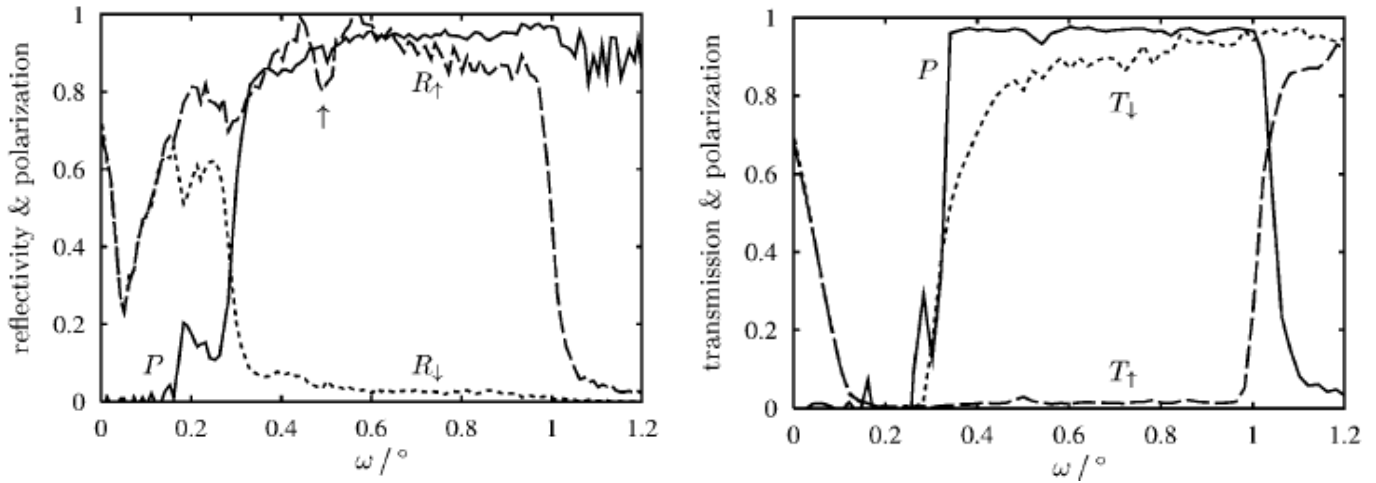

**Fig. II.8.** Reflectivity (left) and transmittance (right) curves for both spin components of the beam, as well as polarization curves for the *Fe/Si* supermirror ($m = 2$) in a small magnetic field ( < 20 Oe) after magnetizing it in a large field (800 Oe) [10].

Thus, examples of remanent supermirror *CoFe/TiZr* and *Fe/Si* coatings with $m = 2$ and 2.5 are demonstrated. When using *Fe/Si* super mirrors with parameter $m > 2.5$, additional research and development may need to be done to improve the remanent properties of polarizing supermirrors.